\documentclass[osajnl,showpacs,superscriptaddress,10pt]{revtex4-1}
\usepackage{dcolumn}% Align table columns on decimal point
\usepackage{bm}% bold math
\usepackage{multirow,array}
\usepackage{graphicx}
\usepackage{amsmath}
\usepackage{array,booktabs}
\usepackage{float}
\usepackage{xcolor}
\usepackage{empheq}
\usepackage{amsmath,amssymb,graphicx}
\begin{document}
\title{Bound states spectrum of the nonlinear Schr\"odinger equation with P\"oschl-Teller  and square potential wells}

\author{L. Al Sakkaf and U. Al Khawaja}\email{Corresponding author: u.alkhawaja@uaeu.ac.ae}
\affiliation{Physics Department, United Arab Emirates University,
P.O. Box 15551, Al-Ain, United Arab Emirates}

\begin{abstract}
We obtain the spectrum of bound states for a modified
P\"oschl-Teller and square potential wells in the nonlinear
Schr\"odinger equation. For a fixed norm of bound states, the
spectrum for both potentials turns out to consist of a finite number
of multi-node localized states. We use modulational instability
analysis to derive the relation that gives the number of possible
localized states and the maximum number of nodes in terms of the
width of the potential. Soliton scattering by these two potentials
confirmed the existence of the localized states which form as
trapped modes. Critical speed for quantum reflection was calculated
using the energies of the trapped modes.
\end{abstract}

\maketitle

\section{Introduction}
\label{introsec} Considerable efforts have been directed towards
understanding the scattering  and interaction dynamics of solitons
with diverse external potentials, for instance, surfaces
\cite{surf1,surf2,surf3}, steps \cite{step1,step2}, potential
barriers \cite{bar0,bar1,bar2,bar3}, potential wells
\cite{well1,well2,well3,well4}, or impurities
\cite{impu1,impu2,impu3,impu4,impu5,impu6}. Various interesting
phenomena occur as a consequence of solitons scattered by
potentials. Quantum reflection is one example that occurs only at
low soliton speeds and demonstrates the wave nature of  solitons. In
such a phenomenon, the soliton is reflected from the potential even
in the absence of a classical turning point
\cite{well1,well2}. Whereas, if the incident soliton velocity is
above a certain critical value, a sharp transition from complete
reflection to complete transmission takes place. This behaviour is
understood as a result of the formation of a localized trapped mode
at the centre of the potential. The resonant interaction between the
incoming soliton and the bound states of the potential well yields
soliton trapping whereas nonlinear interactions initiate the process
of transmission \cite{brand2}. Moreover,  solitons scattered by a
combination of potential wells were exploited to propose a
unidirectional flow of solitons \cite{usa1}, which was then extended
to cases with parity-time symmetric potentials and discrete solitons
in waveguide arrays  \cite{usa4,usa5}.

Scattering of bright solitons by reflectionless potentials, such as
the P\"oschl-Teller (PT) potential, is characterized by the absence
of radiation. The scattering results in either, full transmission,
full reflection, or full trapping \cite{s1}. A sharp transition
occurs at a specific critical speed below which, the soliton fully
reflects and above which it fully transmits the potential well
\cite{well1,s2}. At the critical speed, an unstable trapped mode is
formed where the energy and norm of the incident soliton are equal
to those of the trapped mode at the centre of the potential well
\cite{usa6}. The trapped mode is always formed temporarily during
the scattering process which, for off-resonance scattering, leaves
the potential to join the scattered soliton. An accurate estimate of
the critical speed considering
various potential depths has been provided by Ref. \cite{usa6}.
It has been also shown that, within this setup, a remarkable high-speed
soliton ejection occurs, even for a stationary initial soliton
positioned near the centre of the potential well \cite{usa7}.
Recently, quantum reflection of dark solitons propagating through
potential barriers or in the presence of a
position-dependent dispersion has been also investigated
\cite{dark}.

Identifying the spectrum of bound states is essential to determine
the characteristics of resonant scattering. The spectrum of bound
states and their corresponding energies help in better understanding
the various above-mentioned phenomena. Finding the spectrum will
also provide a physical basis underlying the trapping phenomenon. In
the present study, we consider the NLSE in the presence of the PT
and square (SQ) potential wells. Our primary goal is to obtain the
spectrum of these potentials, namely the profiles and energies of
the bound states. The spectrum of the PT potential well will be
calculated numerically. Motivated by the fact that spectra of
potential wells in general share the same features, we consider the
SQ potential well. The NLSE, in this case is integrable, and can be
solved analytically. The similarities and differences between the
spectra of the PT and SQ potentials will be discussed. We then
investigate the role of bound states on resonant scattering.

The PT potential considered here is modified by relaxing the
reflectionless condition that relates the potential depth, $V_0$,
and inverse width $\alpha$, namely $\alpha=\sqrt{V_0}$. Instead, we
take, $\alpha=\sqrt{V_0}/j$, where $j$ is a nonzero positive
integer. This is motivated by the observations that for the
reflectionless case, $j=1$, only the single-node mode is excited. In
order to be able to excite the multi-nodes trapped modes, it was
necessary to break the reflectionless condition in such a manner.
Remarkably, it turns that the number of nodes in the excited trapped
mode is equal to $j$. Scattering simulation shows that, for $j>1$,
only the trapped mode with the maximum number of nodes forms. We use
a modulational instability (MI) analysis in order to understand and
explain this behaviour. This leads to a formula confirming the
simulation scattering result.

We organize the rest of the paper as follows. In the next section,
we calculate numerically the spectrum of the PT potential. In Sec.
\ref{SQsecA}, we  derive the bound states of the SQ potential.
In Sec. \ref{SQsecB}, we construct the spectrum of bound states and
study its properties. In Sec. \ref{scatsecC}, we calculate the
critical speed for quantum reflection. In Sec. \ref{scatsecA}, we
investigate numerically the scattering dynamics of the bright
soliton by the PT and SQ potential wells. In Sec. \ref{scatsecB}, we
perform the modulational instability analysis to predict the number
of nodes in the trapped mode. Lastly, in Sec. \ref{conc}, we
summarize our findings and conclusions.

\section{Bound states spectrum of P\"oschl-Teller potential well}
\label{PTsec} In this section, we calculate numerically the bound
states for the NLSE with the PT potential well.  The NLSE in
dimensionless form in the presence of an external potential $V(x)$
is written as
\begin{equation}
i\frac{\partial}{\partial t}\psi(x,t)+g_1\frac{\partial^2}{\partial x^2}\psi(x,t)+g_2|
\psi(x,t)|^2\psi(x,t)-V(x)\psi(x,t)=0,
\label{nlse}
\end{equation}
where $\psi(x,t)$ is a complex function and $g_1>0$ and $g_2>0$ are
arbitrary real constants representing the strength of dispersion
and nonlinear terms, respectively. The PT potential we consider here
reads
%\begin{subequation}\label{ptpot}
    \begin{equation}
    \label{ptpot}
    V(x)=-V_0\,{\rm sech}^2(\alpha x),
    \end{equation}
%\end{subequations}
where $V_0>0$ is the the depth of the potential well and
$\alpha=\sqrt{V_0}/j$, being its inverse half width,  $j$ is an arbitrary
nonezero positive integer controlling the potential width. Soliton
scattering becomes reflectionless with $j=1$. The general form of
the stationary state is given by
\begin{equation}
\psi(x,t)=\phi(x) e^{-i \mu t},
\end{equation}
where $\phi(x)$ is a real function and $\mu$ refers to the wave
frequency or, in the case of matter-waves of Bose-Einstein
condensates, it refers to the chemical potential. Substituting in
Eq. (\ref{nlse}) yields the time-independent NLSE
\begin{equation}
\mu\phi(x)+g_1\frac{d^2}{dx^2}\phi(x)+g_2 \phi^3(x)-V(x)\phi(x)=0.
\label{nlse2}
\end{equation}
We are interested in looking for localized symmetric odd parity
solutions defined by  $\phi(-x)=-\phi(x)$. These solutions contain a
node at $x=0$, which implies the initial conditions $\phi_{in}(0)=0$
and $d/dx\phi_{in}(x)|_{x=0}=\delta$, where $\delta$ is an arbitrary
real constant. Another restriction is that the bound state has to
decay to zero outside the potential well, namely ${\rm
lim}_{|x|\rightarrow\infty}\phi(x)\rightarrow0$. This symmetry
reduces the domain of the problem to $[0,\infty]$, which is
sufficient to provide all properties of modes.  There are other
solutions with even parity symmetry defined by $\phi(-x)=\phi(x)$
which do not form a node at $x=0$. In the present work,  we restrict
ourselves to the odd parity solutions since numerical investigations
indicate that scattering a bright soliton by the PT potential always
generates odd parity bound states \cite{usa6, usa7}.  Hence, the
profile of the trapped modes, we are looking for, is composed of
even number of peaks  equally separated by an odd number of nodes
such that there is always a node at the centre of the potential well.

\begin{figure}[!h]
    \centering
    \includegraphics[width=15cm,clip]{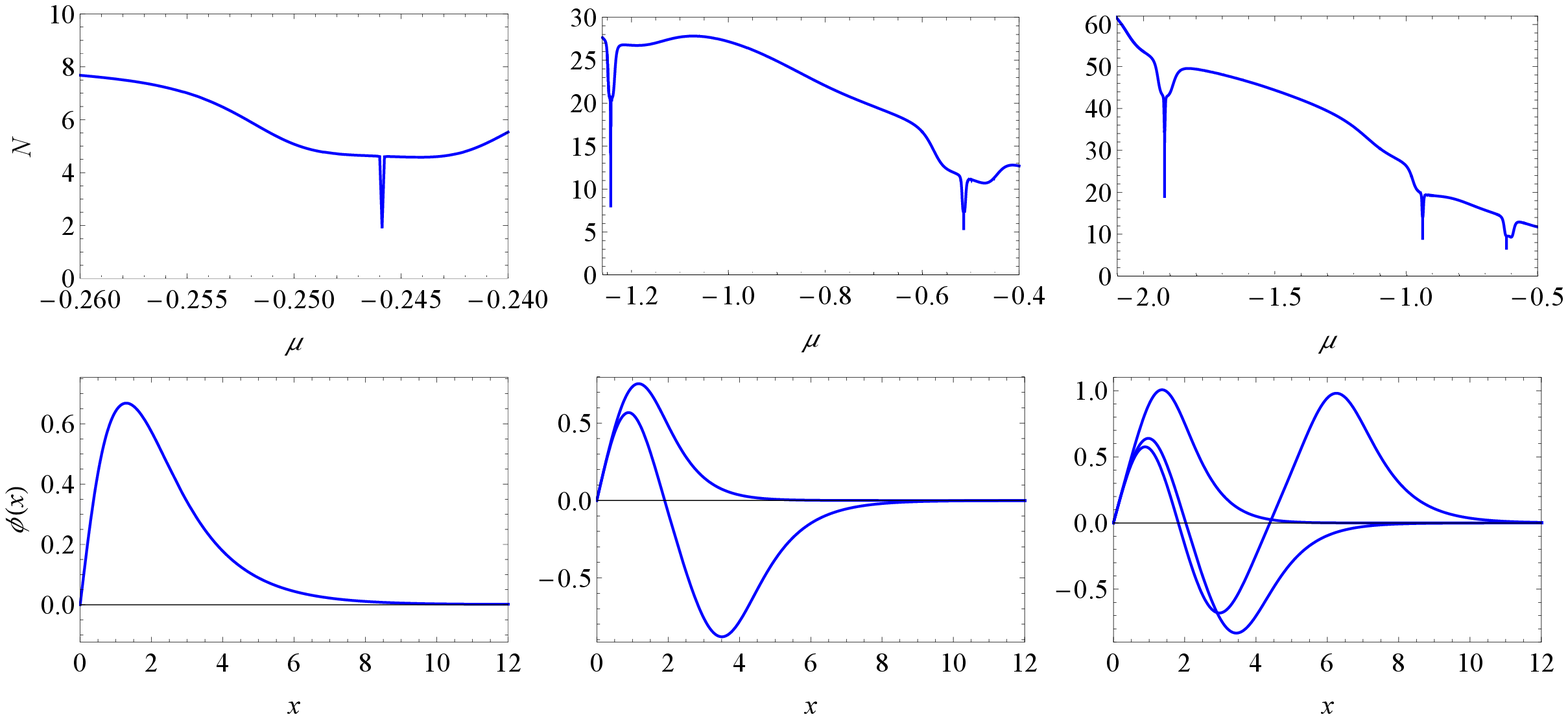}
    \begin{picture}(5,5)(5,5)
    \put(-26,180) {\color{black}{{\fcolorbox{white}{white}{\textbf{(c)}}}}}
    \put(-169,180) {\color{black}{{\fcolorbox{white}{white}{\textbf{(b)}}}}}
        \put(-311,180) {\color{black}{{\fcolorbox{white}{white}{\textbf{(a)}}}}}
    \put(-26,83) {\color{black}{{\fcolorbox{white}{white}{\textbf{(f)}}}}}
        \put(-169,83) {\color{black}{{\fcolorbox{white}{white}{\textbf{(e)}}}}}
        \put(-311,83) {\color{black}{{\fcolorbox{white}{white}{\textbf{(d)}}}}}

    \end{picture}
    \begin{picture}(5,5)(5,5)
    \put(-23,142) {1}
    \put(-46.,149) {2}
    \put(-117,180) {3}
    \put(-82,35) {2}
    \put(-59,87) {1}
    \put(-108,87) {3}

    \put(-172,154) {1}
    \put(-267.6,189) {2}
    \put(-247,53) {1}
    \put(-250,87) {2}
    \end{picture}
    \caption{Norm, as defined by Eq. (\ref{norm0}), in terms of the trapped
soliton frequency (upper row) and the  corresponding profiles of the
possible bound states of each case (lower row). The values of $\mu$
and $N$ of the sharp dips are: (a)
$\mu=-0.245889$, $N=1.8149$ for $j=1$
with bound state profile in (d), (b) $\mu=\{-0.51431,-1.24285\},\,
N=\{5.2209, 7.96884\}$ for $j=3$ with bound state profiles in (e),  and
(c) $\mu=\{-0.618207, -0.937341, -1.91964\}$,  $N=\{6.0509,8.6734,
18.6971\}$ for $j=5$ with bound state profiles in (f).
Parameters used: $g_1=1/2$, $g_2=1$, $\delta=1$, $V_0=2$.}
    \label{pt2}
\end{figure}

\begin{figure}[!h]
    \centering
    \includegraphics[width=7cm,clip]{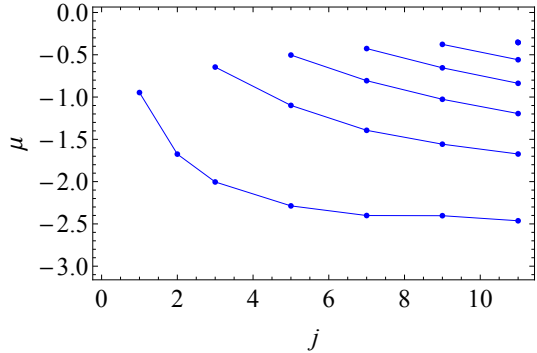}
    \caption{Soliton frequency versus the parameter $j$ for localized solutions that have the
    norm $N=4$. The lines are guides to the eye.
    Lines starting from bottom correspond to the 1-,
3-, 5-, 7-, and 9-node trapped modes, respectively. The single
isolated filled
circle corresponds to the 11-node trapped mode. Parameters used:
$g_1=1/2$, $g_2=1$, $V_0=2$. }
    \label{pt3}
\end{figure}

We start by solving numerically Eq. (\ref{nlse2}) with the
above-mentioned initial and boundary conditions using trial values
of the  soliton frequency $\mu$ and the central slope $\delta$. This
results typically in oscillatory solutions. We then fix the value of
the central slope to a specific value, say $\delta=1$, and start
tuning $\mu$ such that oscillations are pushed out to infinity and a
localized non-oscillatory bound state is obtained. It turns out that
this can be achieved generally with more than one value of $\mu$
such that each value of $\mu$ corresponds to an eigenmode of
different number of nodes and different norm.  The norm of the
resulting state is calculated using
\begin{equation}\label{norm0}
N=\int_{-\infty}^{\infty}|\phi(x)|^2 dx.
\end{equation}
By inspection, we find that the localized mode is always associated
with  a significantly lower norm compared with the oscillatory
solutions. Calculating the norm using (\ref{norm0}) for a range of
$\mu$ values, the critical value is distinguished by a sharp dip in
the curve as shown in Fig. \ref{pt2} for $V_0=2$, where in the upper
row of subfigures, we present three cases with $j=\{1,3,5\}$. In the
lower row, we plot the corresponding profiles of the possible bound
states. This gives an indication on the possible bound states for a
given $j$.
Our objective is then to find the possible number of bound states
for a fixed norm which we set to be $N=4$.
%From the dips shown in Fig. \ref{pt2}, we expect that the number of bound states takes the same  number as the dips.
The value of $\delta$ is  now varied and the tuning procedure of
$\mu$ is repeated for finding all the possible  localized solutions
of a certain $j$ such that they all have the same norm. The filled
circles shown in Fig.  \ref{pt3} represent the coordinates of the
localized solutions that all have the norm $N=4$ in terms of the
soliton frequency $\mu$ and  the parameter $j$. The lines are guides
to the eye connecting the localized solutions that share the same
number of nodes. The figure suggests that for a specific $j$, the
number of possible bound states equals $(j+1)/2$ for odd $j$ and
equals $j/2$ for even $j$.

Our numerical investigation of other potential wells has shown that
the main features of their spectra are common. This motivates the
investigation of integrable case of SQ potential well in the next
section.

\section{Bound states spectrum of square potential well}\label{SQsec}
By inspection, we found that, in general, bound states spectra exist
for a wide range of potential wells and share common main features.
This observation will be exploited to understand the main features
of the spectrum of the PT potential well. To that end, we consider in
this section the finite SQ potential well which is an
analytically-solvable model.

\subsection{Analytic profile of bound states}
\label{SQsecA}
 The finite SQ potential is defined as follows
\begin{subequations}\label{sqrpot}
    \begin{empheq}[left={V(x)=}\empheqlbrace]{align}
    &
    0,\,\,\,\,\,\,\quad\qquad |x|>1/\alpha, {\qquad\rm (\textit{outside})}\\\nonumber\\
    &-V_0, \,\,\qquad |x|<1/\alpha, {\qquad\rm (\textit{inside})}
    \end{empheq}
\end{subequations}
where $V_0>0$ is the the depth of the potential well and
$\alpha=\sqrt{V_0}/j$ is its inverse half width,  and $j$ is an
arbitrary nonezero positive integer.
% The analysis followed here is described briefly as follows:
%\begin{enumerate}
% \item  Finding the exact solution of the system as a set of two subsystems, inside and outside $V(x)$ in the form of Jacobi function ${\rm cn}$.
% \item Applying the initial conditions $\phi_{in}(0)=0$ and its first derivative $d/dx\phi_{in}(x)|_{x=0}$.
% \item Applying the boundary conditions
% $\phi_{in}(1/\alpha)=\phi_{out}(1/\alpha)$ and
%$\frac{d}{dx}\phi_{in}(x)|_{1/\alpha}=\frac{d}{dx}\phi_{out}(x)|_{1/\alpha}$.
%\item Applying the Localization condition $m_2=1$.
%\item Determining the eigenenergies $\mu$ for a certain $b_1$.
%\item Obtaining the eigenmodes corresponding to the determined eigenenergies $\mu$.
%\item Calculating the normalization $N=N_{in}+N_{out}$.
%\item Calculating the mode energy $E=E_{in}+E_{out}$.
%\end{enumerate}
For both regions, inside and outside the potential well, the NLSE is
integrable. The solutions we are seeking are oscillatory inside the
potential well and decaying outside. The exact solution of the NLSE
that describes both cases is the cn Jacobi elliptic function. Inside
the potential well, the solution of the NLSE, (\ref{nlse2}), is
denoted by $\phi_{in}(x)$ and outside of the potential well it is
$\phi_{out}(x)$. Each one of these solutions contains 4 unknown
parameters, as described below. The initial and boundary conditions
will be sufficient to determine all unknown parameters. We solve Eq.
(\ref{nlse2}) for each region separately, as follows:
 \\\\
\textit{Inside the potential well:} In this region, where $V(x)=-V_0$,
the solution takes the form
\begin{equation}\label{fin}
\phi_{in}(x)=c_1\,{\rm cn}[b_1(x+{x_0}_1),m_1],
\end{equation}
where $c_1$, $b_1$, ${x_0}_1$, and $m_1$ are real constants to be
determined. Direct substitution in Eq. (\ref{nlse2}), results in the
following two equations
\begin{equation}\label{m1}
m_1=-\frac{\mu-b_1^2g_1+V_0}{2b_1^2g_1},
\end{equation}
\begin{equation}\label{c1}
c_1=p_1\sqrt{\frac{-\mu+b_1^2g_1-V_0}{g_2}}.
\end{equation}
Here, $p_1=\pm1$ and we have used the identities: ${\rm
dn}(x,m)=\sqrt{1-m\,{\rm sn}^2(x,m)}\,$ and
 $\,{\rm sn}(x,m)=\sqrt{1-{\rm cn}^2(x,m)}$. Without loss of generality, we choose here and below
to express all 7 unknown parameters in terms of $b_1$.
\\\\
\textit{Outside of the potential well:} In this region, $V(x)=0$, and
similarly, the solution takes the form
\begin{equation}\label{fout}
\phi_{out}(x)=c_2\,{\rm cn}[b_2(x+{x_0}_2),m_2],
\end{equation}
with new four parameters $c_2$, $b_2$, ${x_0}_2$, and $m_2$, to be
determined. Substituting in Eq. (\ref{nlse2}), we obtain
\begin{equation}\label{m2}
m_2=\frac{1}{2}-\frac{\mu}{2b_2^2g_1},
\end{equation}
\begin{equation}\label{c2}
c_2=p_2\sqrt{\frac{-\mu+b_2^2g_1}{g_2}},
\end{equation}
where, $p_2=\pm1$ is independent from the sign of $p_1$.
\\\\
\textit{Initial condition:} Here, we apply the initial condition
that will determine the unknown parameter ${x_0}_1$. The initial
condition associated with a wave solution depends on the parity of
the solution.  Since we are restricted to the symmetric odd parity
solutions, the two initial conditions will be: $\phi_{in}(0)=0$ and
$d/dx\phi_{in}(x)|_{x=0}$ is arbitrary. Accordingly, we have
\begin{equation}
c_1 {\rm cn}[b_1\,{x_0}_1,m_1]=0.
\end{equation}
Solving for ${x_0}_1$, we get
\begin{equation}
{x_0}_1=\frac{{\rm K}(m_1)}{b_1}\label{x01},
\end{equation}
where ${{\rm K}(m_1)}$ is the complete elliptic integral of the
first kind.
\\\\
\textit{Boundary conditions:} Here, we apply the boundary conditions
that will determine two more unknown parameters. The continuity of
the solution and its first derivative at $x=1/\alpha$, are expressed as
\begin{equation}\label{bc1}
\phi_{in}(1/\alpha)=\phi_{out}(1/\alpha),
\end{equation}
\begin{eqnarray}\label{bc2}
\frac{d}{dx}\phi_{in}(x)\Big{|}_{1/\alpha}=\frac{d}{dx}\phi_{out}(x)\Big{|}_{1/\alpha}.
\end{eqnarray}
Solving (\ref{bc1}) for $x_{02}$ and (\ref{bc2}) for $b_{2}$, we, respectively, get
\begin{equation}
x_{02}=-\frac{1}{\alpha}+\frac{p_3}{b_2}\,{\rm
cn}^{-1}\left(\frac{c_2}{\phi_0},m_2\right)\label{x02},
\end{equation}
\begin{equation}\label{b2}
b_2=\frac{p_2}{\sqrt{g_1}}\left[\mu^2+2\mu \phi_0^2g_2+g_2(2\phi_1^2g_1+\phi_0^4g_2)\right]^{1/4},
\end{equation}
where we have introduced
\begin{equation}\label{phi0}
\phi_0=c_1\,{\rm
cn}\left[b_1\left(\frac{1}{\alpha}+x_{01}\right),m_1\right],
\end{equation}
\begin{equation}\label{phi1}
\phi_1=-b_1 c_1\,{\rm dn}\left[b_1\left(\frac{1}{\alpha}+x_{01}\right),m_1\right] {\rm sn}\left[b_1\left(\frac{1}{\alpha}+x_{01}\right),m_1\right],
\end{equation}
and $p_3=\pm1$ is independent from the signs of $p_1$ and $p_2$.

Up to this point, all unknown parameters are determined in terms of
a single arbitrary parameter, namely $b_1$. It is just our choice to
leave out this parameter as the arbitrary one; it could have been
any other parameter instead. In the following, we impose the
restriction that the solutions have to decay to zero outside the
potential, which is justified by seeking localized states. This
condition will, essentially, determine the last unknown parameter
and the system of 8 unknowns will be fully determined.
 \\\\
\textit{Localization condition:} Here, an additional condition is
introduced for the solution to decay to a zero background. This can
be achieved by setting $m_2=1$, where the outer solution reduces to
$c_2\,{\rm cn}[b_2(x+x_{02}),1]=c_2\,{\rm sech}[b_2(x+x_{02})]$, which satisfies
(\ref{nlse2}) for
\begin{equation}\label{c22}
c_2=p_2\,\sqrt{\frac{-2\mu}{g_2}},
\end{equation}
\begin{equation}\label{b22}
b_2=p_2\,\sqrt{\frac{-\mu}{g_1}}.
\end{equation}
Substituting for $b_2$ from (\ref{b2}) in (\ref{m2}) with $m_2=1$,
we get the following transcendental equation for $\mu$
\begin{equation}\label{q}
q(\mu)=\mu+\sqrt{b_1^4 g_1^2-V_0(V_0+2\mu)+2V_0(-b_1^2g_1+V_0+\mu)\,{\rm cn}^2\left[b_1\left(\frac{1}{\alpha}+x_{01}\right),m_1\right]}=0.
\end{equation}
The roots of this equation give the eigenfrequencies of localized
modes. It is also noticed that Eqs. (\ref{c22}) and (\ref{b22})
restrict $\mu$ to be negative. Furthermore, we will show next that
real-valued solution profiles are obtained only for values of $\mu$
between two limits, which we denote by $\mu_1$ and $\mu_2$. The
limit $\mu_1$ is defined by the maximum value of $\mu$ for which the
quantity under the square root in Eq. (\ref{c1}) is positive and
thus $c_1$ is real. Setting $c_1=0$ in (\ref{c1}),  gives
\begin{equation}\label{mu1}
\mu_1=b_1^2g_1-V_0.
\end{equation}
%This is a solution to Eq. (\ref{q}) for all $0<b_1\le
%\sqrt{V_0/g_1}$. However, $\mu=\mu_1$ will not be included since it
%results in a trivial solution.
This equation defines a threshold value on $b_1$, namely
${b_1}_{th}=\sqrt{V_0/g_1}$, for which $\mu_1=0$. For $\mu\le\mu_2$,
the quantity $x_{01}$ given by (\ref{x01}) diverges or becomes
complex. Therefore, the value of $\mu_2$ is easily obtained from Eq.
(\ref{m1}) with $m_1=1$, which gives
 \begin{equation}\label{mu2}
 \mu_2=-b_1^2g_1-V_0.
 \end{equation}
To conclude, only roots of $q(\mu)$ located within the interval
 $(\mu_2,\mu_1)$ lead to solutions with real-valued profiles.
 In case $b_1\ge {b_1}_{th}$, Eq. (\ref{mu1}) shows that $\mu_1>0$.
 Since acceptable roots require $\mu<0$,
 then the interval becomes $(\mu_2,0)$.
 \\\\
 \textit{Normalization:} The total normalization $N$ is the sum of the inner
 and outer norms, $N_{in}$ and $N_{out}$, respectively.
 Analytically, it is given by
 \begin{eqnarray}
 N(\mu)=N_{in}+N_{out}&=&2\int_{0}^{1/\alpha}|\phi_{in}(x)|^2dx+2\int_{1/\alpha}^{\infty}|\phi_{out}(x)|^2dx,
 \nonumber\\&=&\frac{2c_1^2}{b_1 m_1 \alpha}\left(b_1(m_1-1)-\alpha\,{\rm E}(m_1)+\alpha\,{\rm E}
 \left\{{\rm am}\left[\frac{b_1}{\alpha}+{\rm K}(m_1),m_1\right],m_1\right\}\right)\nonumber\\&&+\frac{2c_2^2}{b_2}\left\{1-{\rm tanh}\left[b_2\left(\frac{1}{\alpha}+x_{02}\right)\right]\right\}.\label{norm}
 \end{eqnarray}
The second and third lines in the last equation  correspond to
$N_{in}$ and $N_{out}$, respectively,  ${\rm E}(\cdot)$ is the
elliptic integral of the second kind and ${\rm am}(\cdot)$ is the
amplitude of the Jacobi elliptic function. The prefactor $2$ in
front of the integrals accounts for  the complete domain
$[-\infty,\infty]$.

\subsection{Constructing the spectrum}
\label{SQsecB} Since the norm is a conserved quantity, we aim at
constructing the spectrum of bound states for a fixed norm. Taking
into account Eqs. (\ref{m1},\,\ref{c1},\,\ref{x01},\,\ref{phi0},\,\ref{phi1},\,\ref{c22},\,\ref{b22}), the
transcendental Eqs. (\ref{q}) and (\ref{norm}) will be given in
terms of only $b_1$ and $\mu$. In order to obtain a spectrum with
fixed norm, we need to set a value of $N(\mu)$ in (\ref{norm}) and
then solve the system (\ref{q}) and (\ref{norm}) for $b_1$ and
$\mu$. The resulting roots of (\ref{q}) will give the
eigenfrequencies of the bound states. However, this procedure turns
out to be not practical since it requires solving two coupled
transcendental equations simultaneously. Alternatively, we follow
the following approach. We consider a range of $b_1$ values, compute
the roots of $q(\mu)$ for each value of $b_1$, calculate the
associated norm for each root, and then extract from this collection
of data the roots which have the same norm.
 \\

\begin{figure}[!h]
    \centering
    \includegraphics[width=15cm,clip]{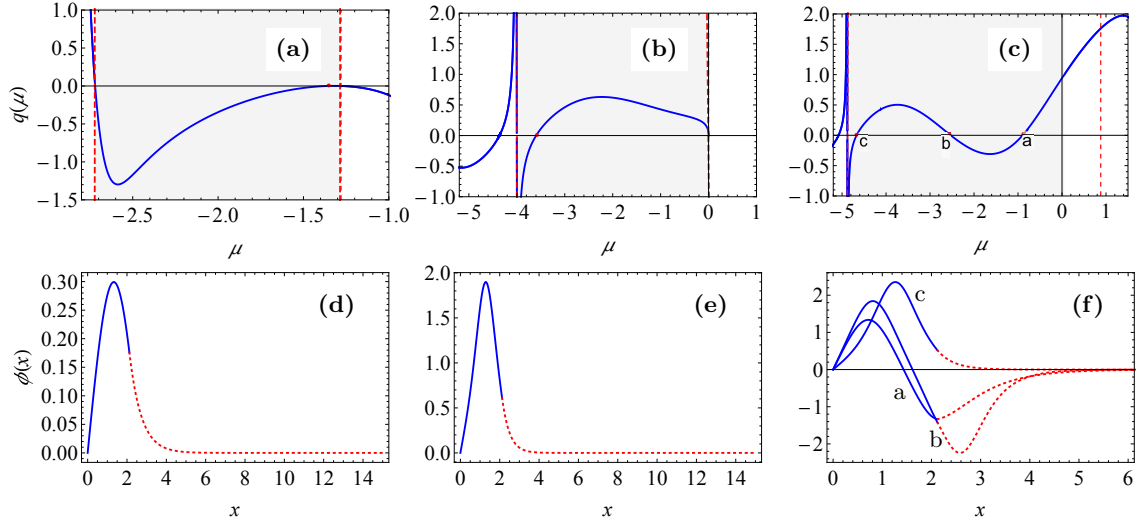}
    \begin{picture}(5,5)(5,5)
    \put(-54,180) {\color{black}{{\fcolorbox{white}{white}{\textbf{(c)}}}}}
    \put(-189,180) {\color{black}{{\fcolorbox{white}{white}{\textbf{(b)}}}}}
    \put(-327,180) {\color{black}{{\fcolorbox{white}{white}{\textbf{(a)}}}}}
    \put(-26,83) {\color{black}{{\fcolorbox{white}{white}{\textbf{(f)}}}}}
    \put(-169,83) {\color{black}{{\fcolorbox{white}{white}{\textbf{(e)}}}}}
    \put(-311,83) {\color{black}{{\fcolorbox{white}{white}{\textbf{(d)}}}}}

    \end{picture}
    \begin{picture}(5,5)(5,5)
    \put(-84.5,32) {b}
    \put(-98,50) {a}
    \put(-90,87) {c}
    \end{picture}
    \caption{Eiegen frequencies and eigen modes for three cases of $b_1$.
    The upper row shows the real roots of Eq. (\ref{q})
    and the lower row presents the corresponding wave profile,
    (a) and (d) for  $b_1<{b_{1}}_{th}$  $(=0.6\,{b_{1}}_{th})$,
    with a single root at $\mu=-1.3695$,
    (b) and (e) for $b_1={b_{1}}_{th}$  with a single root at $\mu=-3.6025$,
    and (c) and (f) for $b_1>{b_{1}}_{th}$ $(=1.2\,{b_{1}}_{th})$,
    with three roots at $\mu=\{-0.9075,-2.5135,-4.6585\}$.
    Filled (red) circles  indicate roots, dashed (red)
    vertical lines indicate the positions of $\mu_1$ (right)
    and $\mu_2$ (left), and the area of included $\mu$ is shaded.
    Parameters used: $g_1=1/2$, $g_2=1$, $V_0=2$, $j=3$.}
    \label{fig1}
\end{figure}

In Fig. \ref{fig1}, we clarify the procedure just described for an
example of SQ potential well characterized by $V_0=2$ and $j=3$. The
left column of the figure corresponds to the case $b_1<{b_1}_{th}$.
The shaded area is used to identify the range of acceptable roots
limited by $\mu_1>\mu>\mu_2$. The limits $\mu_1$ and $\mu_2$ are
indicated by two vertical dashed (red) lines. One eigenfrequency is
found at $\mu=-1.3695$ corresponding to a single-node trapped mode.
The case of $b_1={b_1}_{th}$ is shown in the  middle column of
subfigures. The roots range of this case is $0>\mu>\mu_2$, where
$\mu_1=0$. Similarly, only one root at $\mu=-3.6025$ is found. The
profile of the corresponding single-node trapped mode is different
than the previous one. The last case, shown in the right column of
subfigures, is for $b_1>{b_1}_{th}$. Here, three roots appear at
$\mu=\{-0.9075, -2.5135, -4.6585\}$ and labeled by $\{a, b, c\}$,
respectively. As we mentioned above, the acceptable roots of this
case are in the range $0>\mu>\mu_2$. Interestingly, both $a$ and $b$
wave frequencies form two distinguishable triple-node trapped modes.
An important difference between them should be noted. The mode
corresponding to root $a$ has constant maximum amplitude in the
oscillatory part. For the mode corresponding to root $b$, the peaks
at both edges are larger than those in between. We denote the two
cases by symmetric and asymmetric profiles, respectively. It should
be also noted that the three single-node trapped modes in the three
cases are all different from each other.
The desired smooth and continuous matching between the inner and
outer solutions is attained with $p_1=p_2=1$. The required
sign of $p_3$ is obtained by matching the slopes of the inner and
outer solutions at the edge of the potential. This can be done by
equating $\phi_1$ from Eq. (\ref{phi1}) to a similar expression but
with subscripts $1$ being replaced by $2$.

\begin{figure}
    \centering
    \includegraphics[width=7cm,clip]{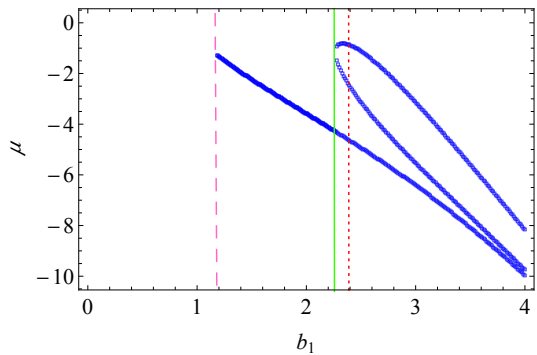}
    \caption{Mode frequency, $\mu$,  in terms of $b_1$. The curves from the bottom of the figure correspond to single-, symmetric triple-, and  asymmetric triple-node trapped modes, respectively. The dashed (pink) line defines the range $b_1\le1.1632$ for which no eigenmodes exist. The solid (green) line at $b_1=2.25169$ indicates the starting value of  triple-node trapped mode formation with the presences of single-node trapped  mode. The  dotted (red) line at $b_1=2.4$ is an example capturing three modes from the bottom of the figure at $\mu=\{-0.9075,-2.5135,-4.6585\}$. Parameters used:  $g_1=1/2$, $g_2=1$, $V_0=2$, $j=3$, $0.5\,{b_1}_{th}\le b_1\le2\,{b_1}_{th}$.}
    \label{fig2}
\end{figure}

Constructing the spectrum, starts, as we described above, with
finding the roots of (\ref{q}) for a range of $b_1$ values. The
result is shown in Fig. \ref{fig2}. The figure shows three curves
which correspond, starting from the bottom, to single-node trapped
modes and two triple-node trapped modes.  No roots exist for
$b_1<1.1632$, which is  indicated by the vertical dashed (pink)
line. A single value of $\mu$ is obtained for a  wide range of
$b_1$, varying from 1.1632 to 2.25169, which is indicated by the
region in between the vertical dashed (pink)  and solid (green)
lines. Three values of $ \mu$ are found for $b_1>2.25169$. As an
example, we draw a vertical dotted (red) line at  $b_1=2.4$ that
crosses the three roots shown in Fig. \ref{fig1}(c). Their
corresponding profiles are those shown in Fig. \ref{fig1}(f).

\begin{figure}
    \centering
    \includegraphics[width=7cm,clip]{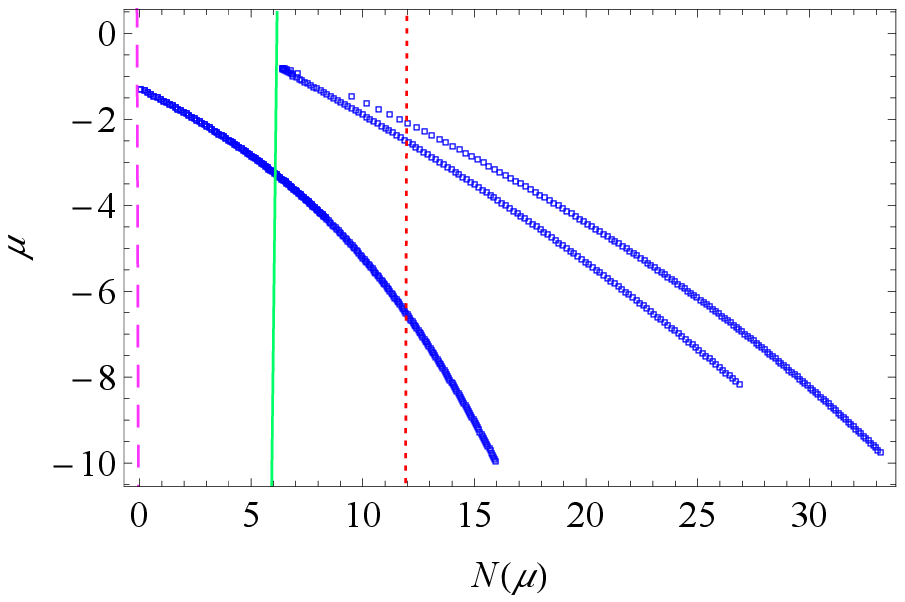}
            \begin{picture}(5,5)(5,5)
    \put(-22,119) {\color{black}{{\fcolorbox{white}{white}{\textbf{(a)}}}}}
    \end{picture}
    \hspace*{0.05cm}
    \includegraphics[width=6.8cm,clip]{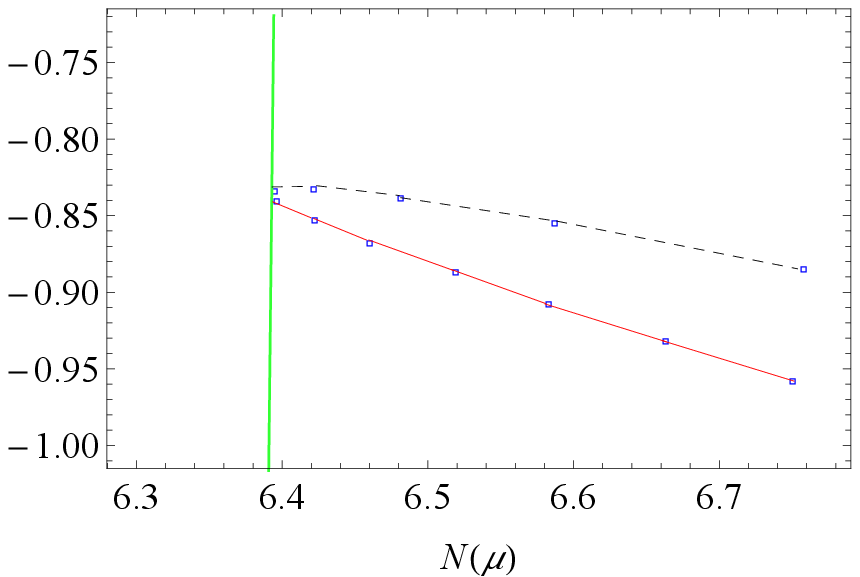}
            \begin{picture}(5,5)(5,5)
    \put(-22,119) {\color{black}{{\fcolorbox{white}{white}{\textbf{(b)}}}}}
    \end{picture}
    \caption{Possible wave frequencies $\mu$ in terms of  normalization $N(\mu)$. The curves from the bottom of the figure (a) correspond to the same curves in Fig. \ref{fig2}. The dashed (pink) line detects the  value of $N=0.06972$ at which no $\mu$ is detected before. The solid (green) line at $N=6.3916$ indicates the minimum value that supports the formation of triple-node mode with the presences of single-node mode. The (red) dotted line at $N=12$ is an example supporting three modes at $\mu=\{-5.7785,-4.6035,-2.2035\}$  from the bottom of the figure, respectively. A zoom-in portion of the plot in (a) is shown in (b) where dashed (black) curve corresponds to asymmetric triple-node modes and solid red curve corresponds to symmetric triple-node modes. Parameters used are the same of those in Fig. \ref{fig2}.}
    \label{fig4}
\end{figure}

Calculating the norm of modes corresponding to all points in Fig.
\ref{fig2}, the relation between $N(\mu)$ and $\mu$ of the three
curves can be extracted and is plotted in Fig. \ref{fig4}(a).
Interestingly, only single-node trapped modes are observed to  occur
with the norm range  $N< 6.3916$. Larger norm is needed for the
formation of higher nodes trapped modes. Two distinguishable triple-node
trapped modes are formed for $N> 6.3916$.  The  dotted (red) line
crosses three roots corresponding, starting from the lowest curve,
to a single-, asymmetric triple-, and  symmetric triple-node trapped
modes,  all with the same norm, $N\approx12$. Figure \ref{fig4}(b)
shows a zoom of the  point at which the two branches of triple-node
trapped modes appear.  The upper branch connected by the dashed
(black) line corresponds to symmetric triple-node trapped modes
while the lower branch connected by the solid (red) line corresponds
to the asymmetric triple-node trapped modes.
\\

% \begin{figure}
%   \centering
%   \includegraphics[width=7cm,clip]{fig6a.eps}
%           \begin{picture}(5,5)(5,5)
%   \put(-42,119) {\color{black}{{\fcolorbox{white}{white}{\textbf{(a)}}}}}
%   \end{picture}
%   \hspace*{0.05cm}
%   \includegraphics[width=7cm,clip]{fig6b.eps}
%           \begin{picture}(5,5)(5,5)
%   \put(-22,119) {\color{black}{{\fcolorbox{white}{white}{\textbf{(b)}}}}}
%   \end{picture}
%   \caption{ Roots and corresponding modes captured by the dotted (red) line in Fig. \ref{fig2}. The three roots of $q(\mu)$ in Eq. (\ref{q}) denoted by $\{a,\, b,\, c\}$ at $\mu=\{-2.2035, -4.6035, -5.7785\}$, respectively, for $b_1=2.278881$ (a).  The corresponding wave profiles are shown in (b). Parameters used: $g_1=1/2$, $g_2=1$, $V_0=2$, $\alpha=\sqrt{V_0}/3$.}
%   \label{fig3}
%\end{figure}
\begin{figure}[h]
    \centering
    \includegraphics[width=7cm,clip]{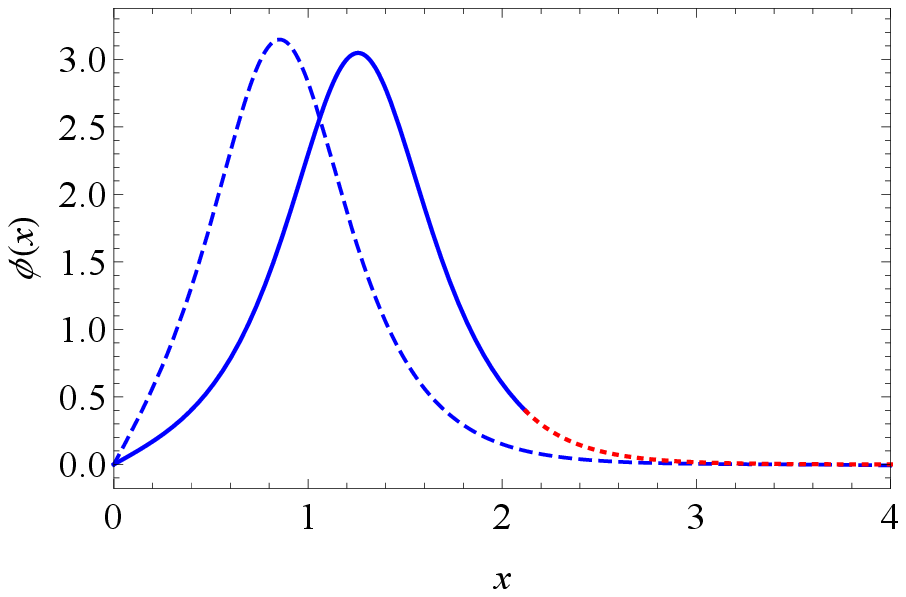}
    \includegraphics[width=6.6cm,clip]{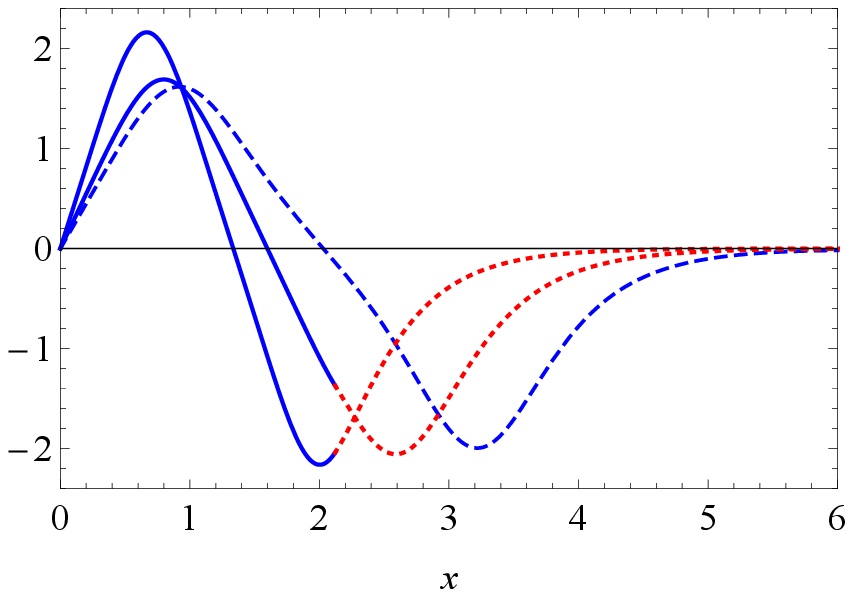}
    \begin{picture}(5,5)(5,5)
    \put(-215,119) {\color{black}{{\fcolorbox{white}{white}{\textbf{(a)}}}}}
    \end{picture}
    \begin{picture}(5,5)(5,5)
    \put(-32,119) {\color{black}{{\fcolorbox{white}{white}{\textbf{(b)}}}}}
    \end{picture}
    \caption{Profiles of the single-node (a) and triple-node (b) trapped modes of the SQ and PT potential
    wells.
        Dashed curves correspond to PT potential well and solid (blue) and dotted (red)
        curves correspond to the SQ potential well.
        Parameters used:  $g_1=1/2$, $g_2=1$, $V_0=2$, $j=3$.}
    \label{fig11}
\end{figure}

\begin{figure}
    \centering
    \includegraphics[width=14.9cm,clip]{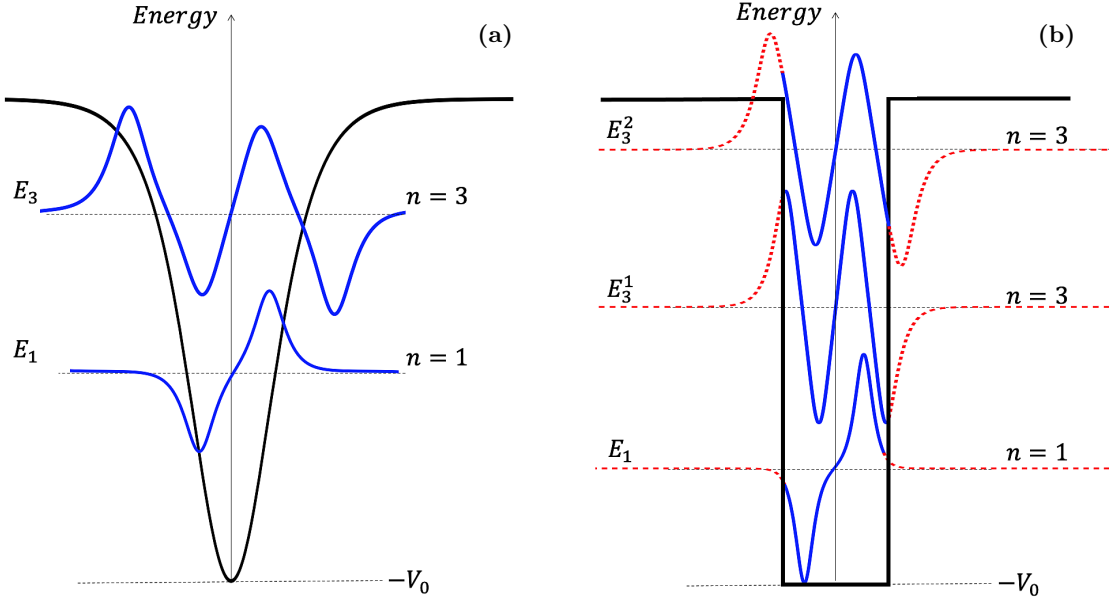}
%    \qquad
%        \includegraphics[width=8.3cm,clip]{fig10a}
            \begin{picture}(5,5)(5,5)
        \put(-28,218) {\color{black}{{\fcolorbox{white}{white}{\textbf{(b)}}}}}
        \put(-240,218) {\color{black}{{\fcolorbox{white}{white}{\textbf{(a)}}}}}
        \end{picture}
    \caption{Schematic diagrams of the spectra  of bound states for: (a) PT potential well, (\ref{sqrpot}), with $E_1$ and $E_3$ corresponding to
    the trapped energies of the
        single- and triple-node trapped modes, respectively,  and (b) SQ potential well, (\ref{ptpot}), with  $E_1, E_3^1,$
        and   $E_3^2$ corresponding to  the trapped energies of the single-, symmetric triple-, and asymmetric
triple-nodes trapped modes the PT potential well. Parameters used are those of Fig. \ref{fig11}.}
    \label{fig7}
\end{figure}

\begin{figure}
    \centering
    \includegraphics[width=7cm,clip]{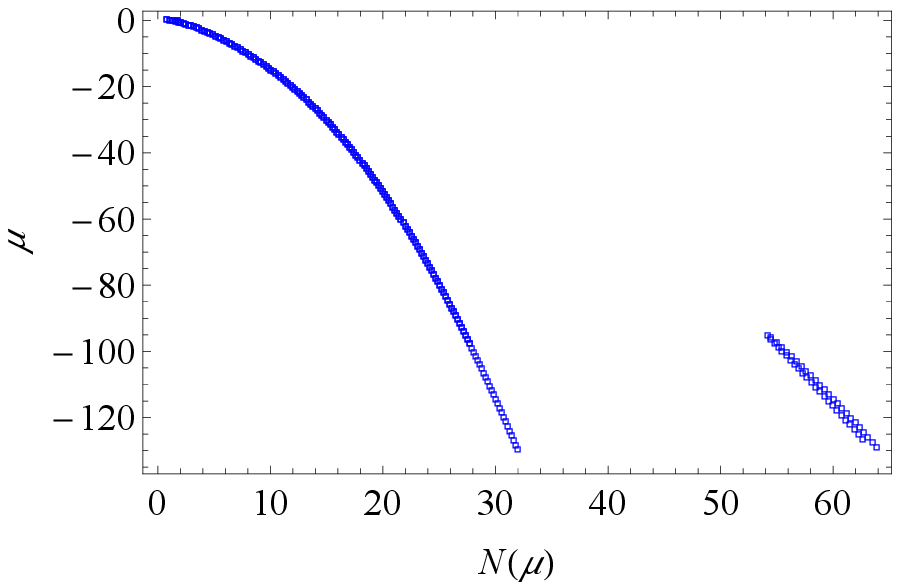}
    \begin{picture}(5,5)(5,5)
    \put(-22,119) {\color{black}{{\fcolorbox{white}{white}{\textbf{(a)}}}}}
    \end{picture}
    \includegraphics[width=6.77cm,clip]{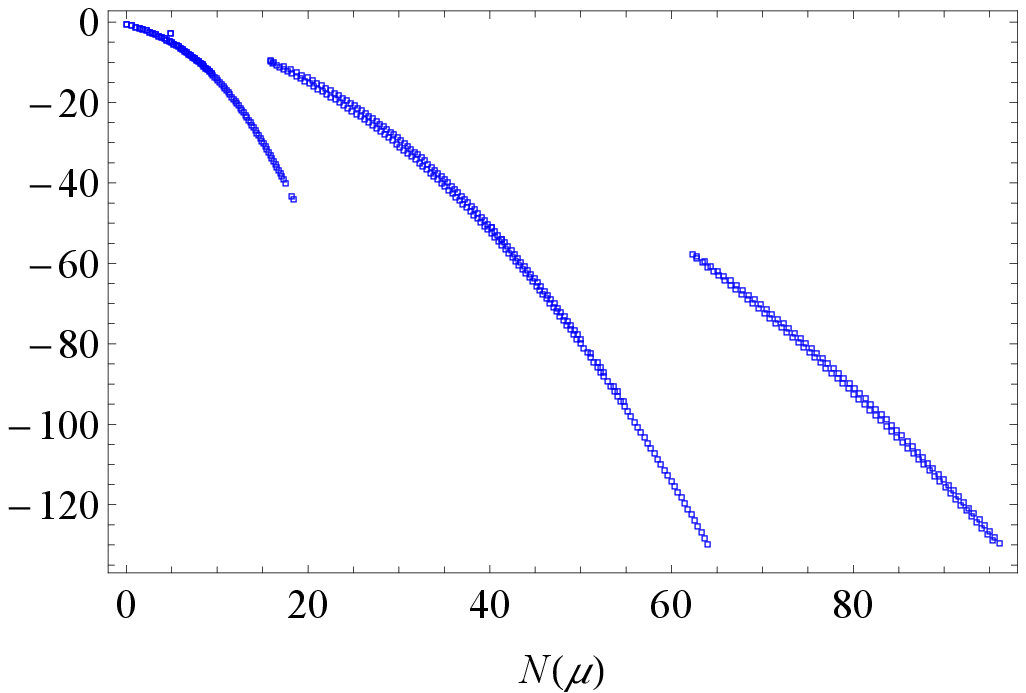}
    \begin{picture}(5,5)(5,5)
    \put(-22,119) {\color{black}{{\fcolorbox{white}{white}{\textbf{(b)}}}}}
    \end{picture}
    \caption{Possible eigenfrequencies in terms of  normalization $N(\mu)$
    for (a) $j=1$ with $0.9\,{b_1}_{th}\le b_1\le8\,{b_1}_{th}$ and (b) $j=2$ with $0.7\,{b_1}_{th}\le b_1\le8\,{b_1}_{th}$.
    The left curve in (a) corresponds to single-node trapped modes while the right curve is composed of two curves corresponding
    to triple-node trapped  modes. In (b), the left curve corresponds to single-node modes, the middle curve is
    composed of two curves both correspond to two triple-node trapped modes, and the right curve is also composed of two
    curves both correspond to quintuple-node trapped modes.  Parameters used:
    $g_1=1/2$, $g_2=1$, $V_0=2$.}
    \label{fig8}
\end{figure}

For the sake of comparison, we show in Fig. \ref{fig11}(a), the
single-node trapped mode profiles of the PT and SQ potential wells
using the same set of parameters.  The profiles of the triple-node
trapped modes are shown as well in Fig. \ref{fig11}(b).  While there
are three trapped modes formed in the SQ potential spectrum, only
two modes are formed in the PT potential spectrum. Inspection
shows that the spectrum of the SQ potential well is always composed
by a finite number of bound states determined by the norm $N$ and
the value of $j$. As $N$ increases, the number of bound states with
the same $j$ increases. However, in the case of the PT potential,
only the parameter $j$ determines the number of the bound states in
the spectrum regardless of the norm. Increasing the norm in the case
of PT potential will not change the number of possible bound states
in the spectrum. Table \ref{teng} summarizes the norm, $N$, energy,
$E_T$, number of nodes, $n$, critical speed for quantum reflection,
$v_c$ (defined in Sec. \ref{scatsecC}), of bound states for the two
spectra. In Fig. \ref{fig7}(a), we show a schematic diagram of  the spectrum for the PT
potential, that is constructed by a single-node and
 triple-node trapped modes with trapped mode energies $E_1$ and $E_3$.
Figure  \ref{fig7}(b), shows as well  the schematic diagram of the spectrum for the SQ
potential that consists of  a single-, symmetric-, and
asymmetric-node trapped modes with trapped mode energies $E_1,
E_3^1,$ and $E_3^2$, respectively.
\\
\begin{table}[H]
    \centering
    \begin{math}
    \begin{array}{c|c|c|c|c}
    %\begin{tabular}{|cccc|}
    \hline
    \multicolumn{5}{c}{\rm\quad PT \,\,potential\,\,}\\\hline
    \mu&n&N&E_T&v_c \\
    \hline
    $-6.4660$ & 1 &12.0465&$-37.3273$&$2.42816$ \\
    \hline
    $-2.22655$& 3^* &12.0270&$-12.3972$&3.16110\\
    \hline
    \multicolumn{5}{c}{\rm\quad SQ \,\,potential\,\,}\\
    \hline
    $-6.6085$ & 1 &12.0469&$-42.0890$&2.25976 \\
    \hline
    $-2.5575$& 3 &12.0362&$-11.0984$&3.19818\\
    \hline
    $-2.1155$ & 3^* & 12.0271&$-9.62501$&3.23320 \\
    \hline
    \end{array}
    %\end{tabular}
    \end{math}
    \caption{Norm, $N$, energy, $E_T$, number of nodes, $n$,
        critical speed, $v_c$, of bound states for
        the SQ and PT potential wells with parameters corresponding to Fig. \ref{fig11}. The 3-node trapped modes distinguished by
        $*$ correspond to the asymmetric modes.}\label{teng}
\end{table}
We show in Fig. \ref{fig8} other cases including the reflectionless
potential with $j=1$, and for a case with $j=2$. A significant
difference between the two potential wells should be noted. While
the reflectionless PT potential does not support other than
single-node trapped modes \cite{usa1}, the SQ potential, supports in
addition to the single-node trapped mode, multi-node trapped modes.
In fact, with the SQ potential well characterized by
$\alpha=\sqrt{V_0}$, multi-node trapped modes do form. The figure
shows up to triple-node trapped modes. With the same maximum value
in the $b_1$ range for the case of $\alpha=\sqrt{V_0}/2$, the
triple-node trapped modes are observed to occur earlier than what
was in the former situation. Moreover, two branches of
quintuple-node trapped modes start to form.
\newpage

\subsection{Critical speed for quantum reflection}
\label{scatsecC} 
\begin{figure}
    \centering
    \includegraphics[width=7cm,clip]{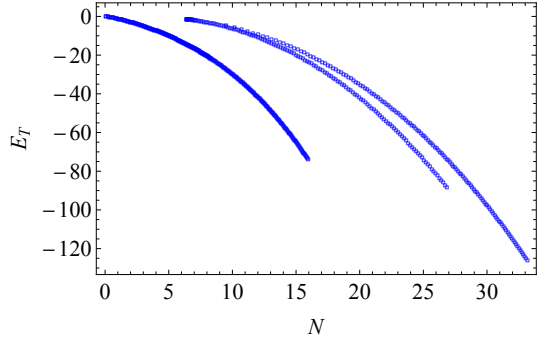}
    \caption{Energy of trapped modes $E_T$ in terms of the norm $N$.
    The curves from the bottom of the figure correspond to the same curves in Fig. \ref{fig2}.
    Parameters used are the same of those in Fig. \ref{fig2}.}
    \label{fig5}
    \end{figure}

\begin{figure}
    \centering
    \includegraphics[width=7.15cm,clip]{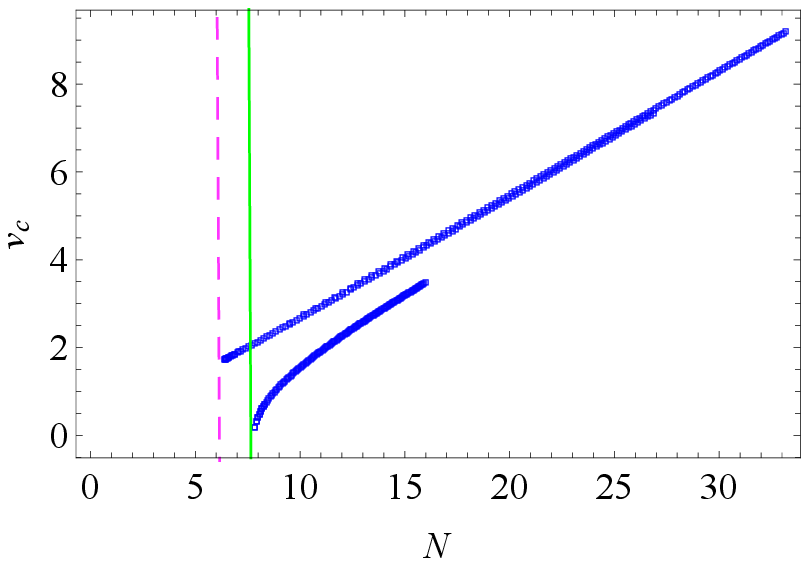}
            \begin{picture}(5,5)(5,5)
    \put(-22,111) {\color{black}{{\fcolorbox{white}{white}{\textbf{(a)}}}}}
    \end{picture}
    \hspace*{0.05cm}
    \includegraphics[width=7cm,clip]{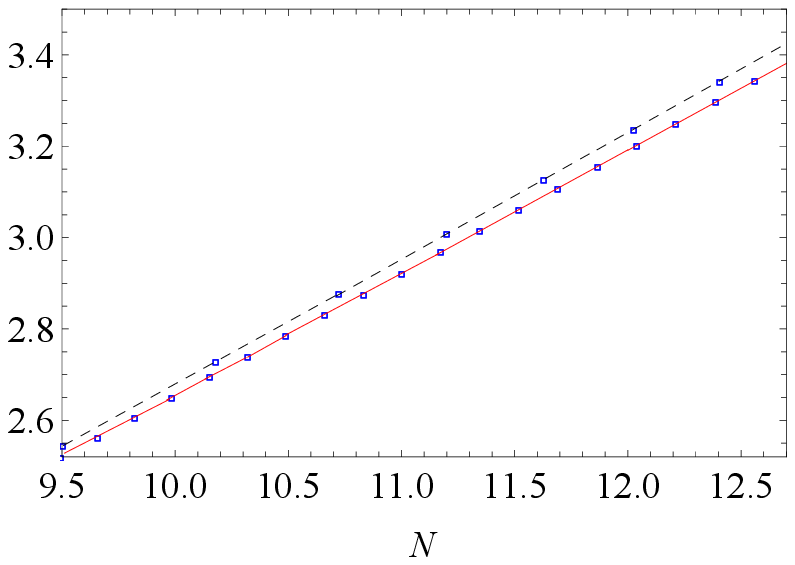}
            \begin{picture}(5,5)(5,5)
    \put(-22,108) {\color{black}{{\fcolorbox{white}{white}{\textbf{(b)}}}}}
    \end{picture}
    \caption{Critical soliton speed, $v_c$, in terms of full normalization $2N$ at which the soliton will be trapped by the potential well (\ref{sqrpot}) forming the modes discussed in Fig. \ref{fig4}. The triple-node modes start to be trapped as a result of soliton quantum reflection with  $N\approx6$, while the single-node trapped modes with $N\approx7.6$.   This is indicated by the solid (green) line. A zoom-in portion of the indistinguishable part in plot (a) is shown in (b) where the dashed (black) curve corresponds to asymmetric triple-node trapped  modes and solid (red) curve corresponds to symmetric triple-node trapped modes. Parameters used are the same of those in Fig. \ref{fig2}.}
    \label{fig6}
\end{figure}

The derivation of the critical speed for quantum
reflection is based on the conservation law of energy. The critical
speed can be obtained by equating the initial energy of the incoming
soliton to that of the trapped mode at the centre of the potential
well. This leads to an analytic formula for the critical speed, as
was shown for the PT potential well \cite {usa1}. Considering the
same scenario for the SQ potential well, the critical speed reads
\begin{eqnarray}\label{cs}
v_c=\sqrt{\frac{1}{12}g_2^2N^2+\frac{2}{N}E_T},
\end{eqnarray}
where $N$ is given by $(\ref{norm})$ and $E_T$ is the energy of the
trapped mode given by
\begin{eqnarray}\label{energy}
E_T=E_{in}+E_{out}&=&N_{in}\,\mu+2\times\frac{g_2}{2}
\int_{0}^{1/\alpha}|\phi_{in}(x)|^4dx+N_{out}\,
\mu+2\times\frac{g_2}{2}\int_{1/\alpha}^{\infty}|\phi_{out}(x)|^4dx
\nonumber\\&=& N\,\mu+\int_{0}^{1/\alpha}|\phi_{in}(x)|^4dx
+\int_{1/\alpha}^{\infty}|\phi_{out}(x)|^4dx.
\end{eqnarray}
Figure \ref{fig5} shows the relation between the trapped mode energy
$E_T$ and the  norm $N$ for the same case of Fig. \ref{fig4}. Using
the same parameters of those in Fig. \ref{fig5}, in Fig. \ref{fig6},
we plot the dependence of the critical speed for quantum reflection
on the trapped mode energy. Three curves are shown in the left
subfigure that correspond to the single- and two triple-node trapped
modes. Although  the upper two curves for the symmetric and
asymmetric triple-node trapped modes seem to have the same $v_c$,
there is a notable difference which is verified by  taking a zoom in
part of the curves, as shown in the right subfigure. The vertical
dashed (pink) line indicates the minimum norm  $N\approx 6$ that is
required for the  quantum reflection to occur with a triple-node
trapped mode. Quantum reflection by the single-node trapped mode is
observed to start at $N\approx 7.6$, as indicated by the solid
(green) line.

\section{Resonant soliton scattering}
\label{scatsec} In this section, we consider the scattering of a
bright soliton by the PT and SQ potential wells in order to confirm
the exitance and the above-described features of the bound states
spectrum. In addition, quantum reflection and its critical speed
values will be also confirmed. This will be followed by a
theoretical proof explaining the specific number of nodes in the
excited trapped modes. This will be based on the modulational
instability analysis of the excited bound state.

\subsection{Scattering dynamics}\label{scatsecA}
\begin{figure}[H]
    \centering
    \includegraphics[width=4.6cm,clip]{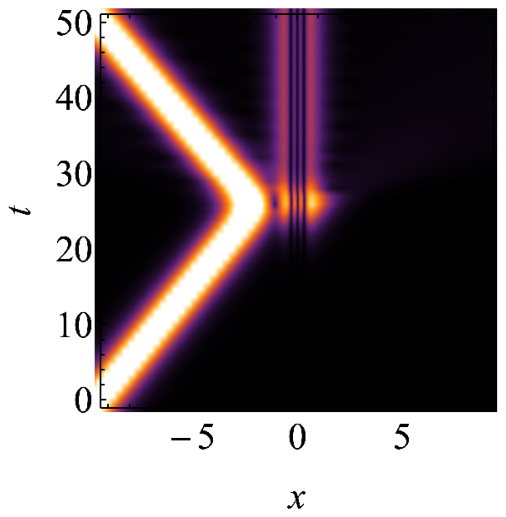}
    \begin{picture}(5,5)(5,5)
    \put(-50,123) {\color{black}{{\fcolorbox{white}{white}{\textbf{(a)}}}}}
    \end{picture}
    \hspace*{0.4cm}
    \includegraphics[width=7cm,clip]{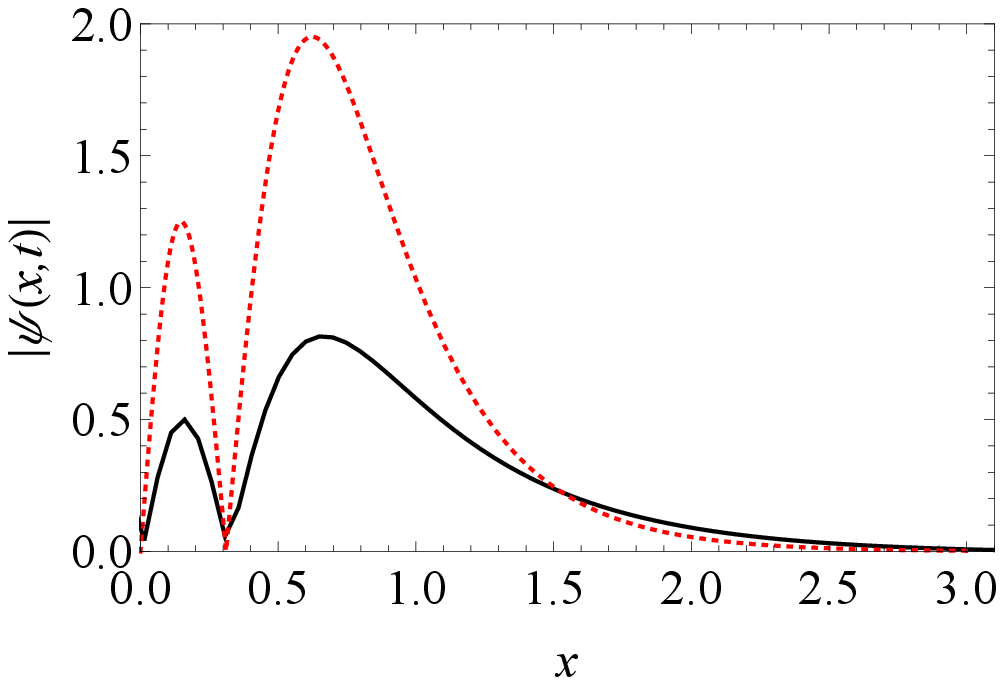}
    \begin{picture}(5,5)(5,5)
    \put(-40,120) {\color{black}{{\fcolorbox{white}{white}{\textbf{(b)}}}}}
    \end{picture}\qquad
        \includegraphics[width=4.6cm,clip]{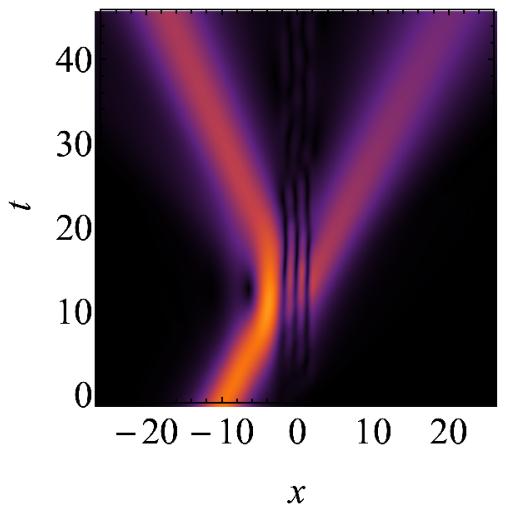}
    \begin{picture}(5,5)(5,5)
    \put(-50,123) {\color{black}{{\fcolorbox{white}{white}{\textbf{(c)}}}}}
    \end{picture}
    \hspace*{0.4cm}
    \includegraphics[width=7cm,clip]{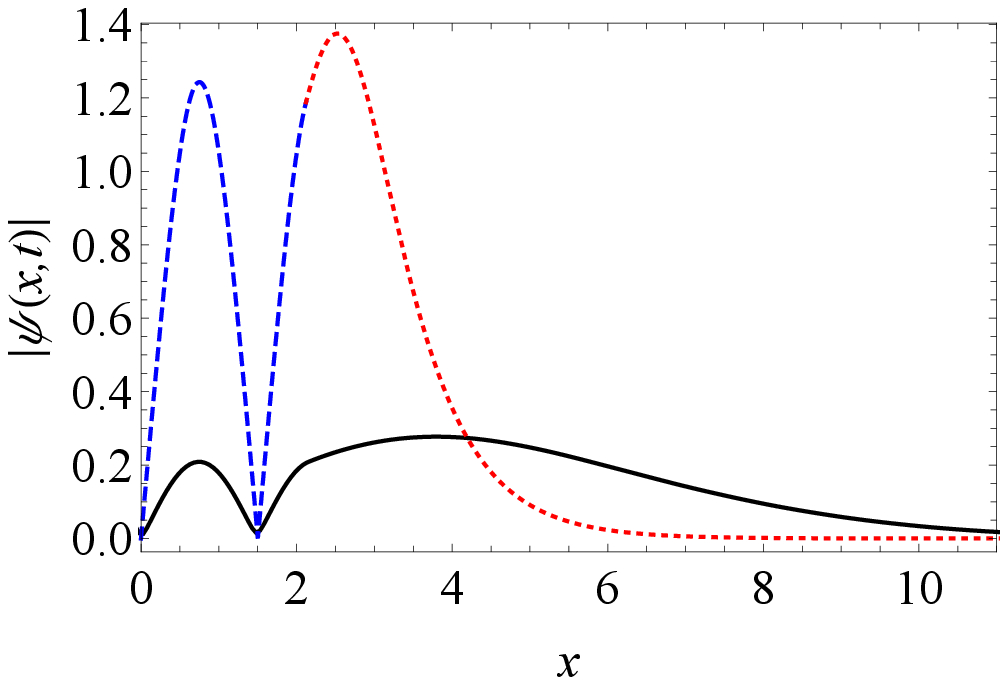}
    \begin{picture}(5,5)(5,5)
    \put(-40,120) {\color{black}{{\fcolorbox{white}{white}{\textbf{(d)}}}}}
    \end{picture}
    \caption{Formation of a triple-node trapped node as a result of the soliton
    described by (\ref{exactsol}) at $t=0$  scattered by (a) the PT potential
    $V(x)=-65\,{\rm sech}^2(\sqrt{65}x/3)$ and $N=4$ and by (c) the SQ potential
    (\ref{sqrpot}) with $V_0=2$ and $N=1$.    (b) Profile of the maximally occupied
    trapped mode  where the  dotted (red) curve is the result of the numerical
    solution of Eq. (\ref{nlse2}) and the solid (black) curve is the maximally
    occupied trapped mode obtained by the scattering shown in subfigure (a).
    (d) Profile of the maximally occupied trapped mode where the
    dashed blue and dotted red curves are the result of the exact solution of Eq.
    (\ref{nlse2}) with $N=7$ and the solid (black) curve is the maximally occupied
    trapped mode obtained by the scattering shown in subfigure (c). Parameters used: $g_1=1/2$, $g_2=1$, $j=3$.}
    \label{fig10}
\end{figure}

Here we describe the scattering setup of a soliton-potential
interaction governed by Eq. (\ref{nlse}). To account for the
theoretical analysis we have made in the previous two sections, we
consider here the scattering of the soliton in two setups; in the
presence of the PT potential well and in the presence of the SQ
potential well, separately.  As an incident soliton, in the two
setups, we use the exact movable bright soliton solution to the
fundamental NLSE, namely, Eq. (\ref{nlse}) with $V(x)=0$, given in a
normalized form as \cite{book}
\begin{equation}\label{exactsol}
\psi(x,t)=N\sqrt{\frac{{g_2}}{8 {g_1}}}\,
\text{sech}\left[\frac{{g_2} N (x-{x_0}-v
    t)}{4 {g_1}}\right] e^{\frac{i}{16 g_1}\left[
    \left({g_2}^2 N^2-4 v^2\right) t+8 v
    \left(x-{x_0}\right)\right]},
\end{equation}
where $x_0$ and $v$ are the initial position and speed of the
soliton centre, and $N$ is its norm given by Eq. (\ref{norm0}).  The
scattering outcome is determined by solving numerically Eq.
(\ref{nlse}) using the iterative power series method \cite{num} with
$\psi(x,0)$ from Eq. (\ref{exactsol}) as an initial profile.
Scattering  dynamics of the soliton described by (\ref{exactsol}) at
$t=0$ with the PT potential $V(x)=-65\,{\rm sech}^2(\sqrt{65}\,x/3)$
and norm $N = 4$ is presented in  Fig. \ref{fig10}(a) where it shows
clearly the formation of a triple-node trapped mode at the centre of
the potential well.  In Fig. \ref{fig10}(b) we plot the
corresponding profiles of the trapped mode obtained by direct
numerical solution of the (\ref{nlse}) and the profile obtained from
the scattering simulation at the classical turning point.  The two
profiles show an agreement on the number and location of the nodes.
However, since the trapped mode is not fully occupied due to
radiation, as the potential is not reflectionless, the maxima in the
profile obtained by the scattering experiment are less than those of
the direct numerical solution. The lower row of the same figure
represents the outcome of the scattering process but with a SQ
potential well characterized by $V_0=2$ and $j=3$.    While the
numerical scattering experiment shows a similar result to that in
the PT potential case,  the formed triple-node mode here is
accompanied with  considerable reflected and transmitted portions,
as shown in Fig. \ref{fig10}(c). Similar to the PT case, the
profiles in Fig. \ref{fig10}(d),  show agreement between the
scattering results and the direct numerical solution on the number
and location of nodes. However, the discrepancy between the
amplitudes of the profiles is larger due to considerable reflection
and transmission.

We have found that, whether the potential is the PT or the SQ
potential well, the scattering process always excites only one bound
state with $j$ number of nodes.  A physical explanation of this
phenomenon will be provided by modulational instability analysis in
the next subsection.

\subsection{Modulational instability analysis}\label{scatsecB}
As discussed above, for each value of $j$, there is a finite number
of eigenmodes. However, only the mode with maximum number of nodes,
that is equal to $j$, is excited by the scattering process. The aim
of this section is to explain this behaviour. To that end, we employ
a MI analysis. Modulational instability is typically caused by
perturbations on a static solution growing up exponentially with
time to either blow up or sometimes form another solution.

We will start by performing the typical MI analysis of a finite
background for the homogeneous NLSE. This will give the frequency of
the most unstable mode. Then, we consider the most unstable mode
within the potential well and apply the boundary conditions to
obtain the relation between the number of nodes and the potential
width.

The so-called constant wave (CW) solution of Eq. (\ref{nlse}) inside
a SQ potential well with depth $V_0$ is given by
\begin{equation}
\psi_0(x,t)=A e^{i (A^2\,g_2+V_0)t},
\end{equation}
where $A$ is an arbitrary real constant. Introducing a small
perturbation $\psi_1(x,t)$ to the CW solution such that
$|\psi_1(x,t)|<<|\psi_0(x,t)|$, we have
\begin{equation}
\psi(x,t)=[A+\psi_1(x,t)] e^{i (A^2\,g_2+V_0)t}.
\end{equation}
Substituting in Eq. (\ref{nlse}) and linearizing in $\psi_1(x,t)$,
we get
\begin{equation}\label{linear}
i\frac{\partial}{\partial t}\psi_1(x,t)+g_1\frac{\partial^2}{\partial x^2}\psi_1(x,t)+A^2\,g_2[\psi_1(x,t)+\psi_1^*(x,t)]=0.
\end{equation}
We assume the perturbation form
\begin{equation}
\psi_1(x,t)=U e^{i(\omega\,t-k\,x)}+V e^{-i(\omega\,t-k\,x)},
\end{equation}
where $k$ and $\omega$ are the wavenumber and frequency of the
perturbation, respectively, and $U$ and $V$ are arbitrary real
constants. Substituting in the linearized equation (\ref{linear}),
results in
\begin{equation}
(A^2g_2)U+(A^2g_2-g_1k^2+\omega)V=0,
\end{equation}
\begin{equation}
(A^2g_2-g_1k^2-\omega)U+(A^2g_2)V=0.
\end{equation}
The condition for nontrivial solution yields the dispersion relation
 \begin{equation}\label{disper}
 \omega=\sqrt{g_1k^2(-2\,A^2\,g_2+g_1\,k^2)}.
 \end{equation}
The imaginary part of $\omega$ has a maximum at
 \begin{equation}\label{wnum1}
  k=\pm A\sqrt{\frac{g_2}{g_1}}.
  \end{equation}
This value corresponds to the most unstable mode. Substituting back
in the dispersion relation (\ref{disper}), the maximum real part of
the frequency is obtained to be $\omega=A^2g_2$. This is the
frequency of the  dominant mode that will determine the fate of the
instability.

Requiring the most unstable CW solution to satisfy the boundary
conditions at the edges of the potential well, defines a specific
wavenumber, which we denote as $k_2$, that depends on the width of
the potential. The CW solution then reads
\begin{equation}\label{MI2}
\psi_0(x,t)=A e^{i(\omega\,t-k_2\,x)}.
\end{equation}
For this CW to be a solution to the NLSE, (\ref{nlse}), the wave
number must satisfy \begin{equation}
k_2=\pm\sqrt{\frac{V_0}{g_1}}\label{k2}.\end{equation} Expressing
$k_2$ in terms of the wavelength of the trapped mode, $\lambda$, as
$k_2=2\pi/\lambda$, Eq. (\ref{k2}) gives
$\lambda=(2\pi\sqrt{g_1})/\sqrt{V_0}$. We define the ratio between
the width of the potential well $W=2/\alpha=2j/\sqrt{V_0}$ and
$\lambda$, as $l=W/\lambda$, which corresponds to the number of
waves inside the potential well. Since each wavelength contributes
with 2 nodes, the number of nodes predicted by MI is given by
$n_{MI}=2l$. We finally obtain the number of nodes
\begin{equation}\label{nodesnum}
n_{MI}=\frac{2}{\pi\sqrt{g_1}}\,j.
\end{equation}
Since the prefactor ${2}/{\pi\sqrt{g_1}}\approx0.9$, the last
formula provides an explaination for the number of nodes in the
trapped modes being equal to $j$. Interestingly, the number of nodes
does not depend on $V_0$.

\begin{figure}[H]
    \centering
    \includegraphics[width=4.5cm,clip]{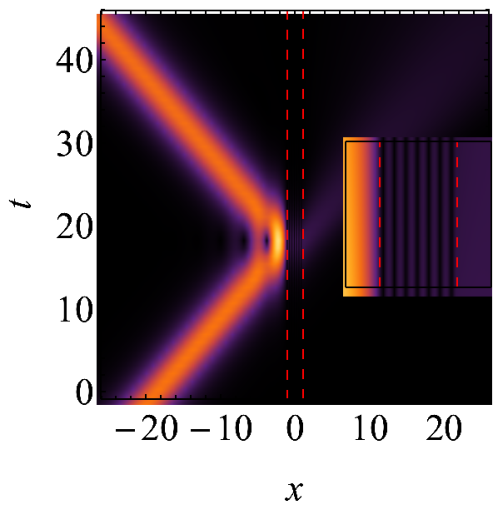}
    \begin{picture}(5,5)(5,5)
    \put(-29,120) {\color{black}{{\fcolorbox{white}{white}{\textbf{(a)}}}}}
    \end{picture}
    \hspace*{0.4cm}
    \includegraphics[width=7.6cm,clip]{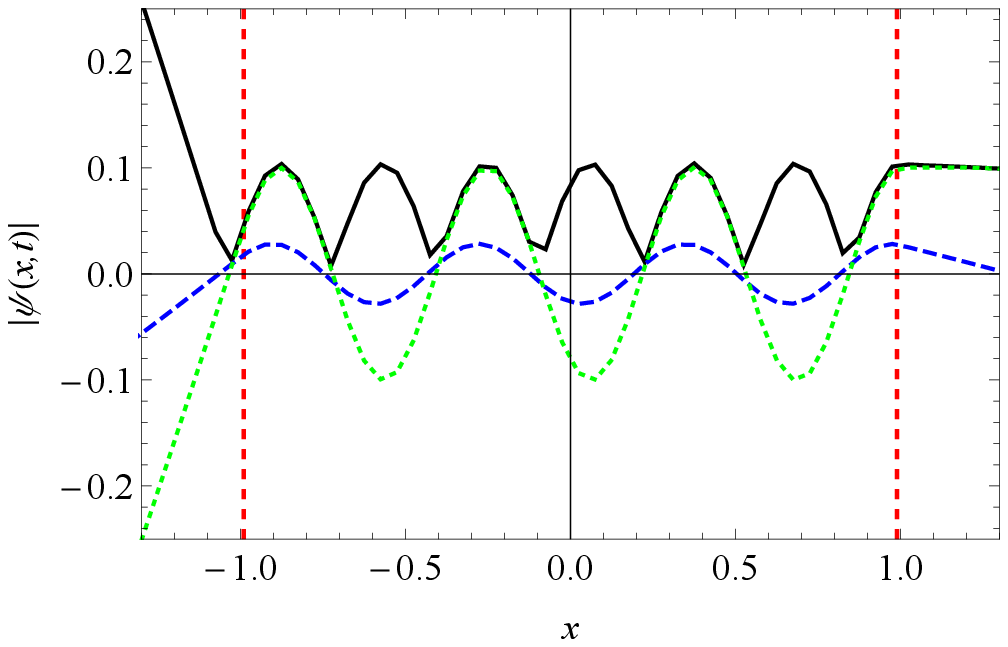}
    \begin{picture}(5,5)(5,5)
    \put(-46,122) {\color{black}{{\fcolorbox{white}{white}{\textbf{(b)}}}}}
    \end{picture}
    \caption{ (a) Formation of trapped node with 7 nodes as a result of  the soliton described by
    (\ref{exactsol}) at $t=0$  scattered by the SQ potential well (\ref{sqrpot}).
    (b) Profile of the maximally occupied trapped mode  where the profile and
    its real and imaginary parts are indicated by solid (black), dashed (blue),
    and dotted (green) curves, respectively. The dashed red vertical lines indicate the
    borders of the potential well width. Inset in (a) shows a zoom-in of the scattering
    near the potential region. Parameters used: $g_1=1/2$, $g_2=1$, $V_0=50$, $j=7$, $N=1$.}
    \label{fig9}
\end{figure}

As an illustrative example,  the formation  of a multi-node trapped
mode for the SQ potential well characterized by $V_0=50$ and $j=7$
and $N=1$ is shown in Fig. \ref{fig9}. The figure shows clearly the
formation of a trapped mode with 7 nodes. The profile of the
maximally occupied trapped mode obtained by the scattering is  also
shown in the left subfigure together with its real and imaginary
parts.

\section{Summary and CONCLUSIONS}\label{conc}
We have revealed the structure of bound states spectrum for a
modified PT potential well. Bound states were obtained through
direct numerical solution of the NLSE, Eq. (\ref{nlse}), using the
potential well (\ref{ptpot}). Tuning the central profile slope and
the frequency, a localized solution with decaying tail is obtained.
The solutions turn out to be characterized by the number of nodes,
their norm, and their energy.  A more efficient alternative method is
to calculate the norm of the numerical solutions for a range of
frequencies where  localized solutions will be identified with sharp
dips in the curve, as shown in Fig. \ref{pt2}.

For a fixed norm, it turns out that a finite number of eigenmodes
exit. Each eigenmode is associated with an eigenfrequency.
Interestingly, the positive nonzero integer $j$, which is used to
define the inverse width of the potential, $\alpha=\sqrt{V_0}/j$,
determines the numbers of possible eigenmodes and their nodes, as follows. For
a given $j$, eigenmodes exist with an odd integer number of nodes,
$n$, such that $1\le n\le j$. The total number of eigenmodes is then
$(j+1)/2$ or $j/2$ for $j$ being odd or even integer, respectively.
This gives the following general structure of the spectrum: it is a
finite number of localized eigenmodes each characterized by its
unique number of nodes and eigenfrequency. Obviously, for $j=1$,
which corresponds to the reflectionless potential case, there is
only one eigenmode which has a single node. This is the well-known
trapped mode responsible for quantum reflection
\cite{well1,impu4,usa6}. It should be noted that the general
structure of the spectrum will not be changed by changing the norm,
$N$; it will affect only the amplitude of eigenmodes profiles.

Motivated by the finding that the general structure of the spectrum
of bound states is a common feature for a wide class of potential
wells, we considered the same problem for the SQ potential well. In
this case, the problem is analytically solvable in terms of the
Jacobi elliptic functions. The spectrum turns out to be, indeed,
similar to that of the PT potential well, but with the difference that
the number of eigenmodes increases with increasing $N$. In addition,
a degeneracy was found where more than one eigenmode having the same
number of nodes, as for instance the two 3-nodes eigenmodes in Fig.
\ref{fig7}, which we denoted as symmetric and asymmetric modes.

Bound states have an important effect on quantum reflection and the
sharp transition in transport coefficients of soliton scattering.
For both potentials, the critical speed for quantum reflection was
calculated using the eigenenergies of the bound states, as
summarized in Table \ref{teng}.

Exciting bound states can be performed by resonant soliton
scattering with the potential at the critical speed for quantum
reflection. Indeed, the numerical experiments show the formation of
trapped modes at the potential with the correct predicted number of
nodes, as shown in Fig. \ref{fig9}. There is no agreement though
between the predicted amplitude and the one obtained from scattering
simulation. This is due to the radiation losses and the fact that in
the SQ potential well, there is a considerable amount of reflected
and transmitted intensities, and thus the trapped mode is not fully
populated.

We have presented a theoretical explanation, based on modulational
instability analysis, for the relation between the predicted number
of nodes, $n_{MI}$, and the integer $j$, as was described above.
This resulted in formula (\ref{nodesnum}) which gives accurately the
predicted number of nodes in terms of $j$.

Numerical simulations of soliton scattering by the potential wells,
for a specific norm $N$ and number $j$, can only  excite the mode
with the maximum number of nodes, namely $j$. Therefore, resonant
scattering by the trapped modes with lower number of nodes, $n<j$,
may not be possible to excite these modes. As an alternative
procedure, we suggest that, exciting such trapped modes
may be achieved through the phase imprinting method, where
the phase extracted from the corresponding analytical solution, is
imprinted initially on a stationary soliton located at the potential
well. This is left to be investigated in a future work.

\section*{acknowledgment}
The authors acknowledge the support of UAE University through grants No. UAEU-UPAR(1)-2019 and No.
UAEU-UPAR(11)-2019.


\begin{thebibliography}{99}
    \bibitem{surf1} F. Baronio, C. De Angelis, P. Pioger, V. Couderc, and A. Barth\'el\'emy, Opt. Lett. {\bf29}, 986 (2004).

    \bibitem{surf2} H. Friedrich and J. Trost, Phys. Rep. {\bf397}, 359 (2004).

    \bibitem{surf3} R. Cote, H. Friedrich, and J. Trost, Phys. Rev. A {\bf56}, 1781 (1997).

    \bibitem{step1} M. Lizunova, O. Gamayun, arXiv:2010.03385 (nlin), (2010).

    \bibitem{step2} Y. Nogami and F.M.Toyama, Phys. Lett. A {\bf184}, 245 (1994).
     \bibitem{bar0} H. Sakaguchi and M. Tamura, J. Phys. Soc. Jpn. {\bf74},  292 (2005).

    \bibitem{bar1} C. Weiss and Y. Castin, Phys. Rev. Lett. {\bf102}, 010403 (2009).

    \bibitem{bar2} O. V. Marchukov, B. A. Malomed, V. A.  Yurovsky, M. Olshanii, V.  Dunjko, and R. G. Hulet,  Phys. Rev. A {\bf99},  063623 (2019).

    \bibitem{bar3} V. Dunjko, M. Olshanii, arXiv:1501.00075v4 (2020).


    \bibitem{well1} C. Lee and J. Brand, Europhys. Lett. {\bf73}, 321 (2006).

    \bibitem{well2} T. Ernst and J. Brand, Phys. Rev. A. {\bf81}, 033614 (2010).



    \bibitem{well3} A. E. Miroshnichenko, S. Flach, and B. Malomed, Chaos {\bf13}, 874 (2003).

    \bibitem{well4} K. T. Stoychev, M. T. Primatarowa, and R. S. Kamburova Phys. Rev. E {\bf70}, 066622 (2004).

    \bibitem{impu1}K. Forinash, M. Peyrard, and B. Malomed, Phys. Rev. E 49, 3400 (1994).

    \bibitem{impu2}X. Cao and B. Malomed, Phys. Lett. A 206, 177 (1995).

    \bibitem{impu3}D. J. Frantzeskakis, G. Theocharis, F. K. Diakonos, P.  Schmelcher, and Y. S. Kivshar, Phys.  Rev. A 66, 053608 (2002).

    \bibitem{impu4}R. H. Goodman, P. J. Holmes, and M. I. Weinstein, Physica D 192, 215 (2004).

    \bibitem{impu5}V. A. Brazhnyi and M. Salerno, Phys. Rev. A 83, 053616 (2011).

    \bibitem{impu6}M. O. D. Alotaibi and L. D. Carr, J. Phys. B, Mol. Opt. Phys. 52, 165301 (2019).


    \bibitem{brand2} T. Ernst and J. Brand,  Phys. Rev. A. {\bf81}, 033614 (2010).

    \bibitem{usa1} M. Asad-uz-zaman and U. Al Khawaja, Europhysics Letters), {\bf101}, 50008 (2013).

%    \bibitem{usa2} U. Al Khawaja, S. M. Al-Marzoug, H. Bahlouli, and Y. S. Kivshar, Phys. Rev. A {\bf88}, 023830 (2013).

%    \bibitem{usa3} M. O. D. Alotaibi, S. M. Al-Marzoug, H. Bahlouli, and U. Al Khawaja, Phys. Rev. E {\bf100}, 042213 (2019).

    \bibitem{usa4} U. Al Khawaja and Andery A. Sukhorukov, Optics Letters, {\bf40}, 2719 (2015).

    \bibitem{usa5} U. Al Khawaja, S. M. Al-Marzoug, and H. Bahlouli, Physics Letters A {\bf384}, 126625 (2020).

    \bibitem{s1}R. H. Goodman, P. J. Holmes, and M. I.Weinstein, Physica D 192, 215 (2004).

    \bibitem{s2}T. Ernst and J. Brand, Phys. Rev. A. 81, 033614 (2010).

    \bibitem{usa6} U. Al Khawaja, Phys. Rev. E {\bf103}, 062202 (2021).

    \bibitem{usa7} T. Uthayakumar, L. Al Sakkaf, and U. Al Khawaja, Phys. Rev. E {\bf104}, 034203 (2021).

    \bibitem{dark} L. Al Sakkaf,  T. Uthayakumar, and U. Al Khawaja, Quantum reflection of dark solitons scattered by reflectionless potential barrier and position-dependent dispersion, Phys. Rev. E (Nov. 2021), submitted.

%    \bibitem{stab1}F. Cooper, A. Khare, A. Comech, B. Mihaila,  J. F. Dawson, and A. Saxena, J. Phys. A: Math. and Theo. {\bf50}, 015301 (2016).
%
%    \bibitem{stab2}J. F. Dawson, F. Cooper, A. Khare, B. Mihaila, E. Arévalo, R. Lan, A. Comech, and A. Saxena, J. Phys. A: Math. and Theo. {\bf50}, 505202 (2017).
%
%    \bibitem{cnoidal1} A. Debnath and A. Khan, The European Phys. J. D \textbf{74}, 1 (2020).
%        \bibitem{s3}N. Kadian, S. Bhatia, S. Pathania, A. Goyal, and N. K. Choragudi, Res. Sqr.  (2021).
%
%    \bibitem{cnoidal2}P. G. Kevrekidis,J. Cuevas–Maraver,A. Saxena, F.  Cooper,and A. Khare,   Phys. Rev. E {\bf92}, 042901 (2015).
%
%    \bibitem{linear1} C. Chen and S. Hu,  Chinese Sci. Abs. Series A \textbf{3}, 29 (1995).
%
%    \bibitem{linear2}J. Lekner, Am. J. Phys. \textbf{875}, 1151 (2007).
%

    \bibitem{book} U. Al Khawaja, and L. Al Sakkaf, {\it Handbook of Exact Solutions to the Nonlinear Schr\"{o}dinger Equations}, IOP Publishing Ltd 2020, London. Online ISBN: 978-0-7503-2428-1, Print ISBN: 978-0-7503-2426-7.

    \bibitem{num} U. Al Khawaja and Q. M. Al-Mdallal, Int. J. Diff. Eq. {\bf2018}, 6043936 (2018); L. Al Sakkaf, Q. M. Al-Mdallal, and U. Al Khawaja, Complexity {\bf2018}, 8269541 (2018);  L. Al Sakkaf and U. Al Khawaja, arXiv:2108.00936

\end{thebibliography}
\end{document}